\documentclass[onecolumn,noshowpacs,superscriptaddress,nobibnotes,nofootinbib,12pt]{revtex4}
\usepackage{amsmath,amssymb}
\usepackage{epsfig}
\usepackage{graphicx}
\newcommand{\be}{\begin{equation}}
\newcommand{\ee}{\end{equation}}
\newcommand{\bea}{\begin{eqnarray}}
\newcommand{\eea}{\end{eqnarray}}
\newcommand{\ben}{\begin{enumerate}}
\newcommand{\een}{\end{enumerate}}
\newcommand{\nn}{\nonumber}

\newcommand{\f}{\frac}

\newcommand{\bra}{\langle}
\newcommand{\ket}{\rangle}

\begin{document}
\author{Bin Wu}
\email{binwu@physik.uni-bielefeld.de}
\affiliation{Frankfurt Institute for Advanced Studies, D-60438 Frankfurt, Germany}
\affiliation{School of Physics and State Key Laboratory of Nuclear Physics and Technology, Peking University, Beijing 100871, China}
\affiliation{Faculty of Physics, University of Bielefeld, D-33501 Bielefeld, Germany}

\title{On $p_{\perp}$-broadening of high energy partons associated with the LPM effect in  a finite-volume QCD medium}
\begin{abstract}
We study the contributions from radiation to $p_{\perp}$-broadening of a high energy parton  traversing a QCD medium with a finite length $L$. The interaction between the parton and the medium is described by decorrelated static multiple scattering.  Amplitudes of medium-induced gluon emission and parton self-energy diagrams are evaluated in the soft gluon limit in the BDMPS formalism. We find both the double-logarithmic correction from incoherent scattering, which is parametrically the same as that in single scattering, and the logarithmic correction from the LPM effect. Therefore, we expect a parametrically large correction from radiation to the medium-induced $p_\perp$-broadening in perturbative QCD.
\end{abstract}

\maketitle

\section{Introduction}
$p_{\perp}$-broadening of high energy quarks and gluons is directly related to the properties of the background medium in relativistic heavy ion collisions. At leading order in perturbative QCD, $p_{\perp}$-broadening in decorrelated multiple scattering takes the following form \cite{Baier:1996sk}
\be
\bra p_\perp^2 \ket_{m0}=\hat{q}L, \label{pt2_BDMPS}
\ee
where $L$ is the length of the medium and the transport coefficient $\hat{q}$ is a characteristic property of the background medium, which can be expressed in terms of the gluon distribution function\cite{Baier:1996sk,Baier:2002tc}. Moreover, in the soft gluon limit  $\bra p_\perp^2 \ket_{m0}$ is found to be closely related to radiative energy loss, that is,\cite{Baier:1996kr,Baier:1996sk,Baier:1998kq}
\be \Delta E \propto \alpha_s N_c \bra p_\perp^2 \ket_{m0} L=\alpha_s N_c \hat{q} L^2.\label{equ:de} \ee 
By determining $\hat{q}$ from experimental data related to $p_\perp$-broadening and radiative energy loss, one can acquire information about the QCD matter created in relativistic heavy ion collisions. 

Leading-order results of $p_{\perp}$-broadening and radiative energy loss in perturbative QCD lead to paradoxical results when confronted with RHIC's data\cite{Eskola:2004cr,Baier:2006fr,Wu:2010ze}.  In Ref. \cite{Wu:2010ze}, it is shown that RHIC's data about $J/\Psi$ suppression indicate that $\hat{q}$ at the thermalization time is less than $1~\mbox{GeV}^2$/fm in most central Au+Au collisions if one uses Equ. (\ref{pt2_BDMPS})\cite{Dominguez:2008be,Dominguez:2008aa}.  In contrast, in Ref. \cite{Eskola:2004cr} the values of the time-averaged  $\hat{q}$ are found to exceed $5~\mbox{GeV}^2$/fm by fitting high-$p_{\perp}$ hadron spectra with radiative energy loss. Besides, such paradoxical results have been already obtained in different analyses about jet quenching(\cite{Baier:2006fr} and references therein). Could such contradictory results be due to that leading order results in perturbative QCD are not so reliable at RHIC energy? To answer this question, one has to calculate corrections of higher orders in $\mathcal{O}(\alpha_s)$ to the above results. In Ref. \cite{CaronHuot:2008}, corrections of $\mathcal{O}(g)$ to $\hat{q}$ have been calculated in an effective theory of QCD. It is found that even at relatively weak coupling $\alpha_s \sim 0.1$ the corrections are already numerically large. 

 On the theoretical side, it is also intriguing to calculate the contributions to $p_\perp$-broadening  from radiation in perturbative QCD, which is of $\mathcal{O}(\alpha_s)$. By parametric analysis, the medium-induced $p_\perp$-broadening from gluon emission in perturbative QCD must have the following form
  \be
 \bra p_\perp^2 \ket_{rad} \simeq \alpha_s N_c C_{rad} Q_s^2, \label{equ:pt2_rad_para}
 \ee
 with $C_{rad}$ a dimensionless coefficient and $ Q_s^2\equiv \hat{q} L$ the saturation momentum squared. In Ref. \cite{Dominguez:2008vd}, it is found that in terms of the appropriate saturation momentum $p_\perp$-broadening in the strongly coupled SYM plasma takes the same parametric form as Equ. (\ref{equ:pt2_rad_para}) if one replaces $\alpha_s N_c$ with $\f{\sqrt{\lambda}}{2\pi}$.  There are no contributions from multiple scattering in SYM plasma. Parametrically, $p_\perp$-broadening is multiple scattering dominated in perturbative QCD and radiation dominated in SYM theory\cite{Dominguez:2008vd}.
 However, the numerical value of $C_{rad}$ has not been evaluated in perturbative QCD in the literature. Therefore, it will help us understand better about these two theories by evaluating $C_{rad}$ to see the limitations of the above parametric conclusion.

In single scattering, the contribution to $p_\perp$-broadening of a high energy quark associated with an uncorrelated gluon emission is double-logarithmically enhanced in the infrared regime\cite{Gunion:1981qs}, that is,
\be
\omega \f{d}{d\omega}\bra p_{R \perp }^2 \ket \simeq \alpha_s N_c \left( \f{L}{\lambda_R} \right)  \bra q^2_\perp \ket_{s}  \int \f{d k_\perp^2}{k_\perp^2} \simeq 2 \alpha_s N_c \hat{q}_R L \log\omega ,\label{equ:idtypeI}
\ee
where $\lambda_R$ is the mean free path of the parton, $\bra q^2_\perp \ket_s \simeq \mu^2$ is the typical transverse momentum squared transferred from the medium to the parton in single scattering, $\mu$ is the screening mass of the constituent particles(scatterers) of the medium and the transport coefficient $\hat{q}_R\sim \f{\mu^2}{\lambda_R}$. Here and in the following, physical quantities of different partons are followed by a subscript $R$ with $R = F$ or $A$ respectively standing for quark jets or gluon jets. Since such a double-logarithmic behavior is well known already, we focus on the contributions to $\bra p_{R \perp }^2 \ket $ from radiation in the LPM regime in this paper.

Our calculations are carried out in the BDMPS formalism\cite{Baier:1996kr,Baier:1996sk,Baier:1998kq}, in which the interaction between high energy partons and the medium is described by decorrelated static multiple scattering. The complete version of the BDMPS formalism in Ref. \cite{Baier:1998kq} gives only details about the calculation of the medium-induced gluon spectrum. We need first to generalize it to the case of the $p_{R\perp}$-differentiated gluon spectrum, that is, $\omega \f{dI_{rg}}{d\omega d^2p_{R\perp}}$. 
Besides, the complete corrections also include the contributions from medium-induced self-energy diagrams. Unfortunately, the evaluation of those diagrams in the LPM regime had not been done in the literature. Therefore, we also need to find a way to calculate the contributions from parton self-energy in the LPM regime inside the medium.

The paper is organized as follows. In Sec. \ref{sec:formalism}, we derive the general formula of the final-state distribution function associated with the LPM effect. We especially focus on the general formula related to the medium-induced self-energy diagrams. The details about the evaluations of all those medium-induced amplitudes defined in Sec. \ref{sec:formalism} are shown in Sec. \ref{sec:amplitudes}. In Sec. \ref{sec:results}, we give the complete results of $p_\perp$-broadening from medium-induced gluon emission diagrams as well as medium-induced parton self-energy diagrams. The physical interpretation of our results is presented in Sec. \ref{sec:discussions}.
%
%
\section{Medium-induced gluon emission and self-energy diagrams}\label{sec:formalism}
When a high energy parton travels in the medium, it can lose its energy due to the medium-induced gluon emission. In the meanwhile the final-state distribution of the parton in transverse momentum space is modified by the gluon recoil. Due to the conservation of probability,  one also needs to consider the contributions from the medium-induced self-energy diagrams, which have not been evaluated in the literature. Therefore, in this section we focus on the general formula for the corrections from medium-induced self-energy diagrams in the BDMPS formalism\cite{Baier:1996kr, Baier:1996sk, Baier:1998kq}. Before that, we first give a brief review on the medium-induced gluon emission.
\subsection{Review of the medium-induced gluon emission}
\begin{figure}
\begin{center}
\includegraphics[width=12cm]{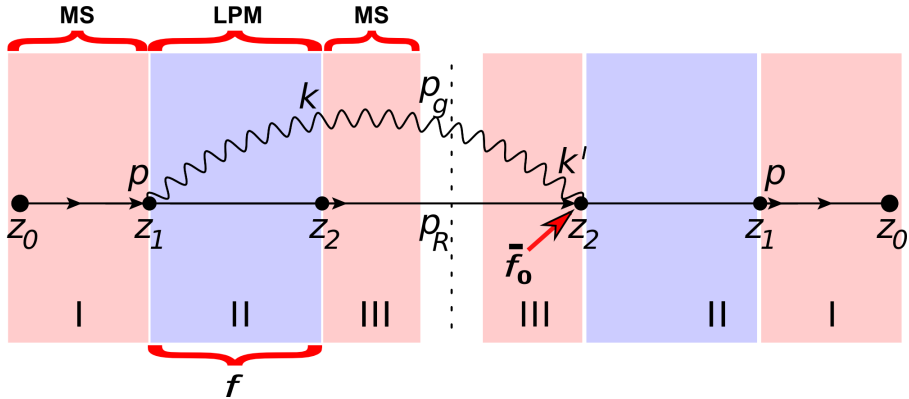}
\end{center}
\caption{Medium-induced gluon emission. In decorrelated multiple scattering, one can separate the whole gluon emission process into three different regions in $z$, the longitudinal coordinate. In region I $(z_1>z>z_0)$, a high energy parton alone is randomly kicked by scatterers.  In region II $(z_2>z>z_1)$, the parton emits a gluon in the amplitude at $z_1$ and later on the emitted gluon is freed from the parton at $z_2$.  In region III $(L>z>z_2)$, the parton is again subject to multiple scattering as a single parton.}\label{fig:ms_rg}
\end{figure}
 Inspired by the discovery of jet quenching at RHIC, radiative energy loss has been extensively studied in the last decade\cite{Baier:1996kr, Baier:1996sk, Baier:1998kq,Zakharov:1996fv, glv, amy,ksw} (see\cite{Wiedemann:2009sh} for a complete review). A typical medium-induced gluon emission process is illustrated in Fig. \ref{fig:ms_rg}. The multiple scattering of the parton-gluon pair in region II is responsible for the so-called LPM effect. In the BDMPS formalism, one denotes the contribution from region II by the full amplitude $\vec{f}$ and the contribution from the last scattering vertex in this region by $\vec{f}_0$. To calculate $p_\perp-$broadening of the leading parton, we introduce the transverse coordinate $x_\perp$ conjugate to the parton's transverse momentum $\vec{p}_{R\perp}$. In the soft gluon limit, the amplitudes $\vec{f}$ and $\vec{f}_0$ depend only on $\vec{x}_\perp$ and $\vec{k}_\perp$, the gluon's transverse momentum. The contributions from multiple scatterings in regions I and III are those from the multiple scattering of a single parton. Putting the contributions in the three regions together, one has the $p_{R\perp}$-differentiated gluon spectrum as follows
\bea
\omega \f{dI_{rg}}{d\omega d^2p_{R\perp}} &=& \f{\alpha_s N_c}{2 } \f{N_c}{C_R} \hat{q}_R^2 \mbox{Re} \int_0^L dz ( L - z )\int  \f{ d^2k_\perp }{(2\pi)^2} d^2 x_\perp e^{-\f{1}{4} \hat{q}_R ( L-z ) x_{\perp}^2 -i ( \vec{p}_{R\perp} + \vec{k}_{\perp}  ) \cdot \vec{x}_{\perp} }\nn\\
&\times& \vec{\bar{f}}_0(\vec{k}_\perp, \vec{x}_\perp) \cdot \vec{f}(\vec{k}_\perp, \vec{x}_\perp,z)\nn\\
&=& \f{\alpha_s N_c}{2 } \f{N_c}{C_R} \hat{q}_R^2 \mbox{Re} \int_0^L dz ( L - z )\int  \f{ d^2B_\perp }{(2\pi)^2} \f{ d^2 x_\perp}{(2\pi)^2} e^{-\f{1}{4} \hat{q}_R ( L-z ) x_{\perp}^2 -i  \vec{p}_{R\perp} \cdot \vec{x}_{\perp} }\nn\\
&\times& \vec{\bar{f}}_0(\vec{B}_\perp-\vec{x}_\perp, \vec{x}_\perp) \cdot \vec{f}(\vec{B}_\perp, \vec{x}_\perp,z), \label{equ:odIrgdodp}
\eea
where $z \equiv z_2 - z_1$, $\vec{B}_\perp$ is the transverse coordinate conjugate to $\vec{k}_\perp$,
\be
\vec{\bar{f}}_0(\vec{B}_\perp, \vec{x}_\perp) \equiv \int d^2k_\perp e^{i\vec{k}_\perp \cdot \vec{B}_\perp}\vec{\bar{f}}_0(\vec{k}_\perp, \vec{x}_\perp),\label{equ:f0bar_def_B}
\ee
and
\be
\vec{f}(\vec{B}_\perp, \vec{x}_\perp,z) \equiv \int d^2k_\perp e^{-i\vec{k}_\perp \cdot \vec{B}_\perp}\vec{f}(\vec{k}_\perp, \vec{x}_\perp,z).\label{equ:f_def_B}
\ee
If one integrates over $\vec{p}_{R\perp}$ and $\vec{x}_\perp$, the above formula reduces to the medium-induced gluon spectrum in Refs. \cite{Baier:1996kr,Baier:1998kq}. 

\subsection{The medium-induced self-energy of high energy partons}

\begin{figure}
\begin{center}
\includegraphics[width=12cm]{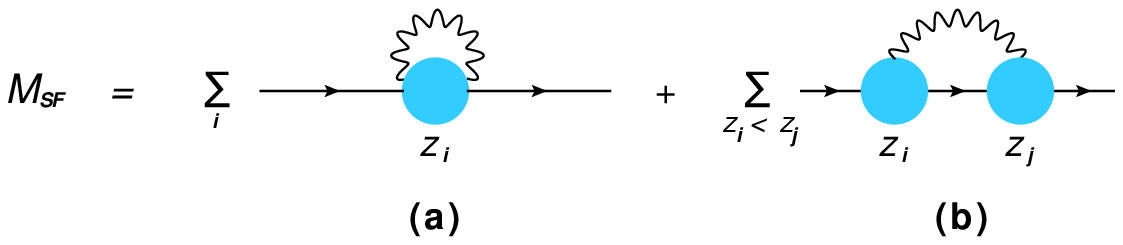}
\end{center}
\caption{Self-energy diagrams in the medium. A process with incoherent gluon emission is illustrated in (a). In this case the virtual gluon is emitted and absorbed within a time $\sim \f{1}{\mu}$ and the corresponding amplitude does not carry the information about the medium through the LPM phases. In self-energy diagrams with coherent gluon emission illustrated in (b), virtual gluons are emitted and absorbed respectively at two different times $z_i$ and $z_j$ with $|z_j - z_i| \gtrsim \lambda_R \gg \f{1}{\mu}$.}\label{fig:mnlo}
\end{figure}
In this subsection, we will derive the general formula for the corrections from medium-induced self-energy diagrams to the distribution function in the BDMPS formalism. As illustrated in Fig. \ref{fig:mnlo}, the screening length $1/\mu$ provides a natural scale that separates physics at long coherent times from that at short times in decorrelated multiple scattering. If the virtual gluon is emitted and absorbed within a time $\sim \f{1}{\mu}$ with only one scatterer involved, the amplitude of the corresponding diagram is the same as that in single scattering.  Since such a virtual gluon does not have enough time to dip into the medium, the amplitude does not carry the information about the medium through the LPM phases. In this section, we only deal with self-energy diagrams with virtual gluons that live much longer than $1/\mu$, which are called {\it medium-induced self-energy diagrams}. In such diagrams, the emitted gluons are allowed only to travel forwards along the positive $z$-direction and one can identify time variables with longitudinal coordinates in the old-fashioned perturbation theory\cite{Baier:1996kr,Baier:1998kq}. 

\begin{figure}
\begin{center}
\includegraphics[width=12cm]{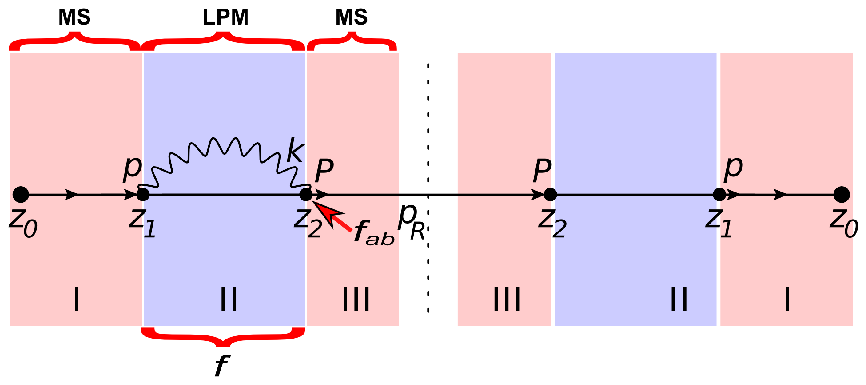}
\end{center}
\caption{The self-energy of a high energy parton in a medium. In decorrelated multiple scattering, one can separate the whole process into three different regions in $z$. In region I $(z_1>z>z_0)$ and region III $(L>z>z_2)$, the parton alone travels in the medium and accumulates its $p_{\perp}$-broadening by random multiple scattering. In region II $(z_2>z>z_1)$, it first emits a gluon at $z_1$; afterwards, the parton-gluon system is subject to multiple scattering before the gluon is absorbed by the leading parton at $z_2$.}\label{fig:ms_vg}
\end{figure}

As illustrated in Fig. \ref{fig:ms_vg}, a medium-induced self-energy diagram is characterized by two time variables: the emission time $z_1$ and the absorption time $z_2$. One can separate the whole process into three different regions in $z$. In region I $(z_1>z>z_0=0)$, a high energy parton is randomly kicked by the scatterers.  In region II $(z_2>z>z_1)$, the parton emits a virtual gluon at $z_1$, which, after being subjected to multiple scattering together with the leading parton, is absorbed at a later time $z_2$. We will account the opposite sequence of times by multiplying a factor of 2 in our final formula. Then, in region III, the leading parton alone continues to accumulate its $p_{\perp}$-broadening from multiple scattering until traveling outside of the medium. Combining the contributions from the three regions, we can write the corrections from medium-induced self-energy diagrams to the distribution function in the following form
\bea
\f{dI_{vg}}{d^2p_{R\perp}} 
&=& \f{\alpha_s N_c}{2 } \f{N_c}{C_R} \hat{q}_R^2 \mbox{Re} \int d^2x_\perp \int_0^L dz (L-z) \nn\\
&\times&   \int \f{ d^2k_\perp}{(2\pi)^2} \f{d\omega}{\omega} \vec{f}_{ab}(\vec{k}_\perp, \vec{x}_\perp)  \cdot   \vec{f}(\vec{k}_\perp, \vec{x}_\perp,z) e^{-\f{1}{4} \hat{q} (L-z) x_\perp^2 -i \vec{p}_{R\perp}  \cdot \vec{x}_\perp} \nn\\
&=& \f{\alpha_s N_c}{2 } \f{N_c}{C_R} \hat{q}_R^2 \mbox{Re} \int \f{d^2x_\perp}{(2\pi)^2} \int_0^L dz (L-z) \nn\\
&\times&   \int \f{ d^2B_\perp}{(2\pi)^2} \f{d\omega }{\omega} \vec{f}_{ab}(\vec{B}_\perp, \vec{x}_\perp)  \cdot  \vec{ f}(\vec{B}_\perp, \vec{x}_\perp,z) e^{-\f{1}{4} \hat{q} (L-z) x_\perp^2 -i \vec{p}_{R\perp}  \cdot \vec{x}_\perp},\label{equ:dIvgdp}
\eea
where $e^{-\f{1}{4} \hat{q} (L-z) x_\perp^2}$ is the contribution from multiple scatterings in regions I and III, $\vec{f}(\vec{k}_\perp, \vec{x}_\perp,z)$ is the full amplitude in the LPM region (region II), which is the same as that used in (\ref{equ:odIrgdodp}), and $ \vec{f}_{ab}$ is the amplitude of the gluon absorption vertex at $z_2$. Here,
 \be
\vec{f}_{ab}(\vec{B}_\perp, \vec{x}_\perp) \equiv \int d^2k_\perp e^{i\vec{k}_\perp \cdot \vec{B}_\perp}\vec{f}_{ab}(\vec{k}_\perp, \vec{x}_\perp).
\ee

In terms of these amplitudes we define the distribution function from radiation as
\be
\f{dI_{rad}}{d^2p_{R\perp}} = \f{dI_{vg}}{d^2p_{R\perp}} + \int d\omega \f{dI_{rg}}{d\omega d^2p_{R\perp}}.\label{equ:dIraddp_def}
\ee
The conservation of probability requires that
\be
\int d^2p_{R\perp} \f{dI_{rad}}{d^2p_{R\perp}} = \int d^2p_{R\perp}\f{dI_{vg}}{d^2p_{R\perp}} + \int d^2p_{R\perp} \int d\omega \f{dI_{rg}}{d\omega d^2p_{R\perp}}=0.\label{equ:probability}
\ee
%
%
\section{Calculations of medium-induced gluon emission/absorption amplitudes}\label{sec:amplitudes}

In this section, we calculate the gluon emission amplitudes $\vec{\bar{f}}_0$, $\vec{f}_{ab}$ and $\vec{f}$ following Ref. \cite{Baier:1998kq}. Among them, evaluations of $\vec{\bar{f}}_0$, $\vec{f}_{ab}$ and $\vec{f}_0\equiv\vec{f}(z=0)$ involve only the gluon emission/absorption in single scattering. In contrast, the calculation of $\vec{f}$ is more involved and can be obtained by solving the Schr\"{o}dinger-like evolution equation in the BDMPS formalism.

\subsection{The medium-induced gluon emission/absorption amplitudes in single scattering}

\begin{figure}
\begin{center}
\includegraphics[width=14cm]{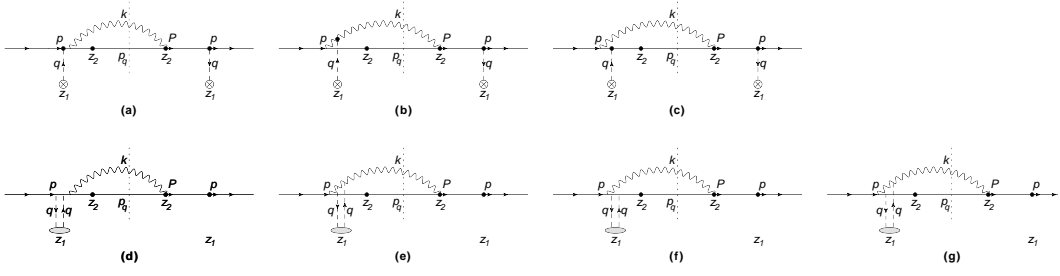}
\end{center}
\caption{Diagrams contributing to $\vec{f}_0$\cite{Baier:1998kq}. Here, one needs to include all the diagrams with gluon emission right before or after $z_1$. Transverse momenta are transferred from the medium to the parton-gluon system in (a), (b) and (c). Therefore, we need to put in a phase factor $e^{ i \vec{q}_{\perp} \cdot \vec{x}_{\perp} }$ in their amplitudes to keep track of the accumulation of transverse momenta.}\label{fig:f0}
\end{figure}

We first evaluate $\vec{f}_0(\vec{B}_\perp, \vec{x}_\perp)\equiv\vec{f}(\vec{B}_\perp, \vec{x}_\perp,0)$. Here, one needs to include all the gluon emission diagrams with only one scatterer at $z_1$ involved. All of such diagrams  are shown in Fig. \ref{fig:f0}, which are the same as those in Ref. \cite{Baier:1998kq}. The color factors and the gluon emission vertices are already known. The difference is that we need to put in a phase factor $e^{ i \vec{q}_{\perp} \cdot \vec{x}_{\perp} }$ with $\vec{x}_{\perp}\equiv \vec{x}_{R\perp} - \vec{x}^\prime_{R\perp}$ in the amplitudes of Fig.s \ref{fig:f0}(a),  \ref{fig:f0}(b) and  \ref{fig:f0}(c) to record the information about the transverse momenta transferred from the medium to the parton-gluon system. In this way, we get the following amplitudes from the gluon emission vertex in each diagram
\bea
&&M_{0a} = - 2g \f{\vec{k}_\perp}{k_\perp^2} e^{i\vec{q}_\perp \cdot \vec{x}_\perp} ,~M_{0b} = 2g \f{\vec{k}_\perp - \vec{q}_\perp}{( k_\perp - \vec{q}_\perp)^2} \left(\f{ N_c }{2 C_R}\right) e^{i\vec{q}_\perp \cdot \vec{x}_\perp},~M_{0c} = 2g \f{\vec{k}_\perp}{k_\perp^2}\left(1 - \f{N_c }{2 C_R}\right)e^{i\vec{q}_\perp \cdot \vec{x}_\perp},\nn\\
&&M_{0d} + M_{0f} = 0,~M_{0e} = 2g \f{\vec{k}_\perp}{k_\perp^2}\left(-\f{N_c}{2 C_R} \right),~M_{0g} = 2g \f{\vec{k}_\perp - \vec{q}_\perp}{( k_\perp - \vec{q}_\perp)^2}  \left( \f{N_c}{2 C_R} \right).\label{equ:f0amp}
\eea
Summing over all the contributions above, we get
\bea
\vec{f}_0(\vec{k}_\perp, \vec{x}_\perp) &=& \left( \f{C_R}{g \pi N_c \hat{q}_R} \right)  \rho  \int d^2q_\perp \f{d \sigma_R}{d^2q_\perp} \sum\limits_{k=a}^g M_{0k} \nn\\
&=& - \f{\rho\sigma_R}{\pi \hat{q}_R} \int d^2q_\perp \f{1}{\sigma_R} \f{d \sigma_R}{d^2q_\perp} \left( 1 + e^{i\vec{q}_\perp \cdot \vec{x}_\perp}\right)\nn\\
&\times&\left[ \f{\vec{k}_\perp}{\vec{k}_\perp^2} - \f{\vec{k}_\perp - \vec{q}_\perp}{ ( \vec{k}_\perp - \vec{q}_\perp)^2} \right],\label{equ:f0k}
\eea
where the differential cross-section is defined as
\be
\f{d\sigma_R}{d^2q_\perp} =\f{\alpha_s
C_R}{\pi}|V(\vec{q}_\perp)|^2.
\ee 
In impact parameter space, it takes the following form
\bea
\vec{f}_0(\vec{B}_\perp, \vec{x}_\perp) &=&  \f{2 \rho\sigma_R i}{ \hat{q}_R} \int d^2q_\perp \f{1}{\sigma_R} \f{d \sigma_R}{d^2q_\perp} \left( 1 + e^{i\vec{q}_\perp \cdot \vec{x}_\perp}\right)\left( 1 - e^{-i\vec{q}_\perp \cdot \vec{B}_\perp}\right) \f{\vec{B}_\perp}{\vec{B}_\perp^2}\nn\\
&\simeq& i \f{\vec{B}_\perp}{\vec{B}_\perp^2} \vec{B}_\perp \cdot ( \vec{B}_\perp- \vec{x}_\perp).\label{equ:f0}
\eea
Here, we have used
\be
\int d^2 k_\perp \f{\vec{k}_\perp}{k_\perp^2} e^{-i\vec{k}_\perp \cdot \vec{B}_\perp} = - 2 \pi i \f{\vec{B}_\perp}{B_\perp^2}.
\ee

\begin{figure}
\begin{center}
\includegraphics[width=14cm]{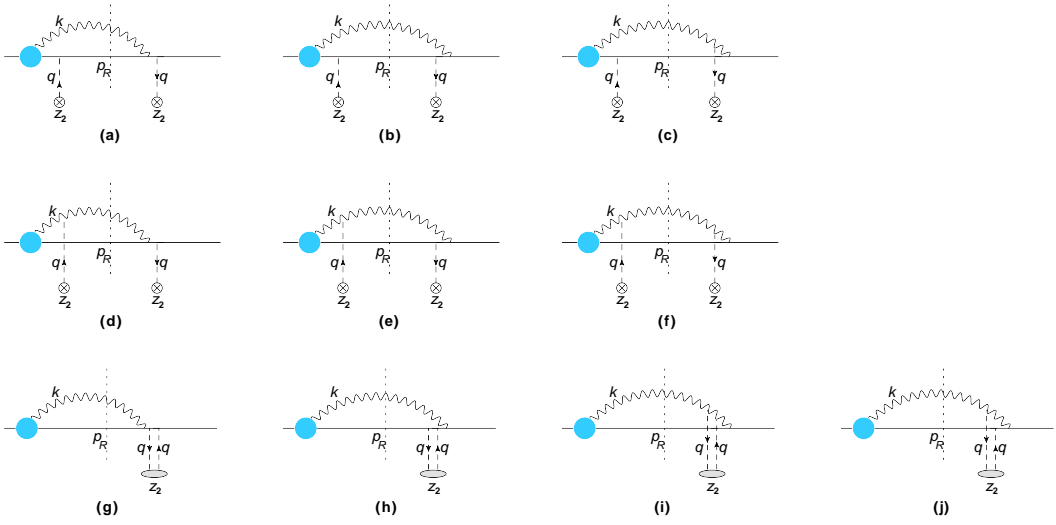}
\end{center}
\caption{Diagrams contributing to $\vec{\bar{f}}_0$\cite{Baier:1998kq}. After being scattered at $z_2$, the parton-gluon system  behaves like a single parton, and the transverse momenta of the emitted gluon will not change any more. Therefore, we should only put in a phase factor $e^{ i \vec{q}_{\perp} \cdot \vec{x}_{\perp} }$ in  (a), (b) and (c), in which transverse momenta are transferred from the medium to the leading parton in the amplitude.}\label{fig:f0bar}
\end{figure}

Next, let us calculate $\vec{\bar{f}}_0(\vec{B}_\perp, \vec{x}_\perp)$. One should take into account all the gluon emission diagrams with only one scatterer at $z_2$ involved in the conjugate amplitude. All the relevant diagrams are shown in Fig. \ref{fig:f0bar}. After being scattered at $z_2$ the parton-gluon system  behaves like a single parton, and the transverse momenta of the emitted gluon will not change any more. Therefore, only those diagrams with transverse momenta transferred from the medium to the leading parton in the amplitude are given a common phase factor $e^{ i \vec{q}_{\perp} \cdot \vec{x}_{\perp} }$.  The amplitude of each diagram in Fig. \ref{fig:f0bar} is as follows
\bea
&&\bar{M}_{0a} =- 2g \left(1 - \f{N_c}{2 C_R}\right)e^{i\vec{q}_\perp \cdot \vec{x}_\perp} \f{\vec{k}_\perp}{k_\perp^2}, ~\bar{M}_{0b} =2g \f{\vec{k}_\perp}{k_\perp^2} e^{i\vec{q}_\perp \cdot \vec{x}_\perp}, ~\bar{M}_{0c} =  2g  \left(-\f{ N_c }{2 C_R}\right)\f{\vec{k}_\perp - \vec{q}_\perp}{( \vec{k}_\perp - \vec{q}_\perp)^2} e^{i\vec{q}_\perp \cdot \vec{x}_\perp},\nn\\
&& ~\bar{M}_{0d} = \bar{M}_{0e} = -\bar{M}_{0j} = 2g  \left(- \f{N_c}{2 C_R} \right) \f{\vec{k}_\perp - \vec{q}_\perp}{( \vec{k}_\perp - \vec{q}_\perp)^2}, ~\bar{M}_{0f} = -2 \bar{M}_{0i} = 2g \f{N_c}{ C_R} \f{\vec{k}_\perp}{k_\perp^2} ,~\bar{M}_{0g} + \bar{M}_{0h}=0.\nn
\eea
As a result, we have
\bea
\vec{\bar{f}}_0(\vec{k}_\perp, \vec{x}_\perp) &=& \left( \f{C_R}{g \pi N_c \hat{q}_R} \right)  \rho  \int d^2q_\perp \f{d \sigma_R}{d^2q_\perp} \sum\limits_{k=a}^j \bar{M}_{0k} \nn\\
&=& \f{\rho\sigma_R}{\pi \hat{q}_R} \int d^2q_\perp \f{1}{\sigma_R} \f{d \sigma_R}{d^2q_\perp} \left( 1 + e^{i\vec{q}_\perp \cdot \vec{x}_\perp}\right)\nn\\
&\times&\left[ \f{\vec{k}_\perp}{\vec{k}_\perp^2} - \f{\vec{k}_\perp - \vec{q}_\perp}{ ( \vec{k}_\perp - \vec{q}_\perp)^2} \right],
\eea
and
\bea
\vec{\bar{f}}_0(\vec{B}_\perp, \vec{x}_\perp)&\simeq& i \f{\vec{B}_\perp}{\vec{B}_\perp^2} \vec{B}_\perp \cdot ( \vec{B}_\perp+ \vec{x}_\perp).\label{equ:f0bar}
\eea
As shown in Fig. \ref{fig:f0bar}, at $z_2 = L$ the diagrams with the gluon emitted inside the medium in the amplitude and emitted outside of the medium in the conjugate amplitude are already accounted in our calculation of $\vec{\bar{f}}$.

\begin{figure}
\begin{center}
\includegraphics[width=14cm]{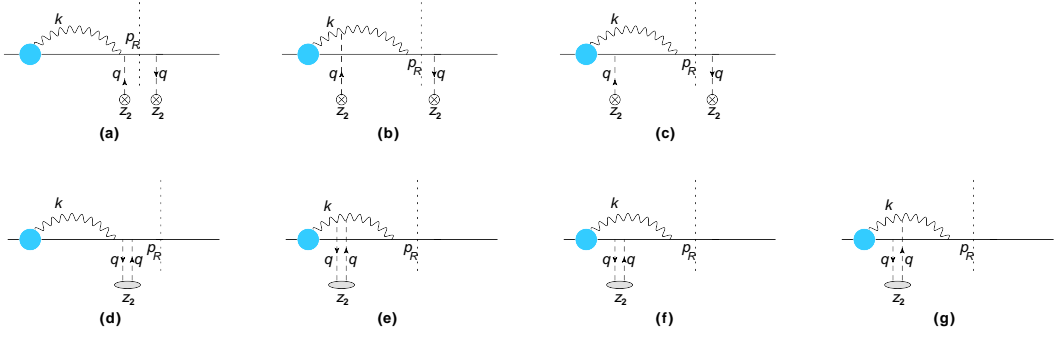}
\end{center}
\caption{Diagrams contributing to $\vec{f}_{ab}$. Each diagram here is the time-reversal of the corresponding diagram in Fig. \ref{fig:f0}. One can simply get $\vec{f}_{ab}$ by replacing $\vec{k}_\perp$ with $-\vec{k}_\perp$ in Equ. (\ref{equ:f0amp}) and multiplying it by -1.}\label{fig:fab}
\end{figure}

As shown in Fig. \ref{fig:fab}, diagrams contributing to $\vec{f}_{ab}$ have a one-to-one correspondence with those contributing to $\vec{f}_0$, whose amplitudes are given by
\bea
&&M_{aba} = - 2g \f{\vec{k}_\perp}{k_\perp^2} e^{i\vec{q}_\perp \cdot \vec{x}_\perp} ,~M_{abb} = 2g \f{\vec{k}_\perp + \vec{q}_\perp}{( k_\perp + \vec{q}_\perp)^2} \left(\f{ N_c }{2 C_R}\right) e^{i\vec{q}_\perp \cdot \vec{x}_\perp},~M_{abc} = 2g \f{\vec{k}_\perp}{k_\perp^2}\left(1 - \f{N_c }{2 C_R}\right)e^{i\vec{q}_\perp \cdot \vec{x}_\perp},\nn\\
&&M_{abd} =- M_{abf} = g \f{\vec{k}_\perp}{k_\perp^2},~M_{abe} = 2g \f{\vec{k}_\perp}{k_\perp^2}\left(-\f{N_c}{2 C_R} \right),~M_{abg} = 2g \f{\vec{k}_\perp + \vec{q}_\perp}{( k_\perp + \vec{q}_\perp)^2}  \left( \f{N_c}{2 C_R} \right).
\eea
Accordingly,
\bea
\vec{f}_{ab}(\vec{k}_\perp, \vec{x}_\perp) &=& - \f{\rho\sigma_R}{\pi \hat{q}_R} \int d^2q_\perp \f{1}{\sigma_R} \f{d \sigma_R}{d^2q_\perp} \left( 1 + e^{i\vec{q}_\perp \cdot \vec{x}_\perp}\right)\nn\\
&\times&\left[ \f{\vec{k}_\perp}{\vec{k}_\perp^2} - \f{\vec{k}_\perp + \vec{q}_\perp}{ ( \vec{k}_\perp + \vec{q}_\perp)^2} \right],
\eea
and
\bea
\vec{f}_{ab}(\vec{B}_\perp)&\simeq& - i \f{\vec{B}_\perp}{\vec{B}_\perp^2} \vec{B}_\perp \cdot ( \vec{B}_\perp- \vec{x}_\perp).\label{equ:fab}
\eea
Taken $z_2 = L$ in Fig. \ref{fig:fab}, one can easily see that the diagrams with the virtual gluon emitted in the medium and absorbed outside of the medium are already accounted in our calculation of $f_{ab}$.

\subsection{The medium-induced gluon emission amplitude $\vec{f}$ at time $z$}

\begin{figure}
\begin{center}
\includegraphics[width=12cm]{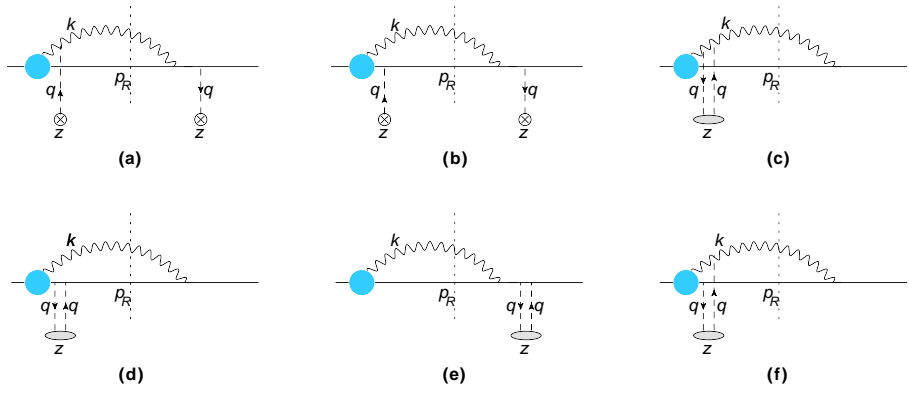}
\end{center}
\caption{Diagrams involving scattering of the parton-gluon system off one scatterer at position $z$. In (a) and (b) there is an extra phase factor  $e^{ i \vec{q}_{\perp} \cdot \vec{x}_{\perp} }$ because transverse momenta are transferred from the medium to the leading particle in their conjugate amplitudes.}\label{fig:fatz}
\end{figure}

To get $\vec{f}$ at any time $z$, we first find the evolution equation. Given $\vec{f}$ at $z$, one can choose an infinitesimal time interval $dz$ such that the
difference between $\vec{f}(z+dz)$ and $\vec{f}(z)$ can be calculated by considering at most one extra scattering. In transverse momentum space we have
\bea
&&\vec{f}(\vec{k}_\perp, \vec{x}_\perp, z+dz) - \vec{f}(\vec{k}_\perp, \vec{x}_\perp, z)  = i\f{k_\perp^2}{ 2\omega } dz\vec{f}(\vec{k}_\perp, \vec{x}_\perp, z)\nn\\
&& + \f{ \rho \sigma_R dz }{C_R} \int d^2 q_\perp \f{1}{\sigma_R} \f{d\sigma_R}{d^2q_\perp}\left[ \f{N_c}{2} e^{i\vec{q}_\perp \cdot \vec{x}_\perp} \vec{f}(\vec{k}_\perp - \vec{q}_\perp, \vec{x}_\perp, z) + \left( C_R - \f{N_c}{2} \right) e^{i\vec{q}_\perp \cdot \vec{x}_\perp} \vec{f}(\vec{k}_\perp , \vec{x}_\perp, z)\right.\nn\\
&&\left. -\f{N_c}{2} \vec{f}(\vec{k}_\perp , \vec{x}_\perp, z) -\f{C_R}{2} \vec{f}(\vec{k}_\perp , \vec{x}_\perp, z) -\f{C_R}{2} \vec{f}(\vec{k}_\perp , \vec{x}_\perp, z) + \f{N_c}{2} \vec{f}(\vec{k}_\perp -\vec{q}_\perp , \vec{x}_\perp, z)  \right] ,
\eea
where the first term on the right hand side results from the free evolution of the parton-gluon system\cite{Baier:1998kq} and the second term comes from the contributions of one extra scattering shown in Fig. \ref{fig:fatz}. There is an extra phase factor  $e^{ i \vec{q}_{\perp}\cdot \vec{x}_{\perp} }$ in the terms corresponding to Fig.s \ref{fig:fatz}(a) and \ref{fig:fatz}(b) because transverse momenta are transferred from the medium to the leading particle in their conjugate amplitudes. In the limit $x_\perp = 0$, the above equation reduces to Equ. (18) in Ref.  \cite{Baier:1998kq} in the soft gluon limit. In impact parameter space, we have
\bea
&&i \f{\partial}{\partial z} \vec{f}(\vec{B}_\perp, \vec{x}_\perp,z) =  \f{1}{2\omega}\nabla_{B}^2 \vec{f}(\vec{B}_\perp, \vec{x}_\perp,z)\nn\\ 
&&+ \f{i}{2} \f{N_c}{C_R} \rho \sigma_R \left\{ V(\vec{B}_\perp) + V(\vec{B}_\perp - \vec{x}_\perp) - 2 - \left( \f{ 2 C_R - N_c }{N_c} \right) \left[ 1 - V(-\vec{x}_\perp) \right] \right\} \vec{f}(\vec{B}_\perp, \vec{x}_\perp,z)\nn\\
&&\simeq \left\{ \f{1}{2\omega}\nabla_{B}^2 -  \f{i N_c\hat{q}_R}{8C_R} \left[ \vec{B}_\perp^2 + (\vec{B}_\perp - \vec{x}_\perp)^2 + \f{ 2 C_R - N_c }{N_c} \vec{x}_\perp^2 \right] \right\} \vec{f}(\vec{B}_\perp, \vec{x}_\perp,z),\label{equ:evotionoff}
\eea
with
\be
V(\vec{B}_\perp) \equiv  \int d^2q_\perp \f{1}{\sigma_R} \f{d \sigma_R}{d^2q_\perp} e^{-i\vec{q}_\perp \cdot \vec{B}_\perp}\simeq 1 - \f{1}{4} \int d^2q_\perp \f{q_\perp^2}{\sigma_R} \f{d \sigma_R}{d^2q_\perp} B_\perp^2.
\ee
After changing variable $\vec{\bar{B}}_\perp = \vec{B}_\perp - \f{\vec{x}_\perp }{2}  $, we have
\bea
&&i \f{\partial}{\partial z} \vec{f}(\vec{\bar{B}}_\perp, \vec{x}_\perp,z) \simeq \left\{ \f{1}{2\omega}\nabla_{\bar{B}}^2 -  \f{i N_c\hat{q}_R}{4C_R} \left[ \vec{\bar{B}}_\perp^2 + \f{ 4 C_R - N_c }{4N_c} \vec{x}_\perp^2 \right] \right\} \vec{f}(\vec{\bar{B}}_\perp, \vec{x}_\perp,z).\label{equ:evolutionf}
\eea
Let us define
\be
\vec{f}(\vec{\bar{B}}_\perp, \vec{x}_\perp,z)= e^{-\f{1}{16} \f{ 4 C_R - N_c }{C_R} \hat{q}_R z \vec{x}_\perp^2 } \vec{\tilde{f}}(\vec{\bar{B}}_\perp, \vec{x}_\perp,z)
\ee
with
\be
  \vec{\tilde{f}}(\vec{\bar{B}}_\perp, \vec{x}_\perp,0) = \vec{f}(\vec{\bar{B}}_\perp, \vec{x}_\perp,0)=i \f{\vec{\bar{B}}_\perp + \f{\vec{x}_\perp}{2}}{ \left( \vec{\bar{B}}_\perp + \f{\vec{x}_\perp}{2} \right)^2}\left( \bar{B}_\perp^2 - \f{x_\perp^2}{4} \right).
\ee
According to (\ref{equ:evolutionf}), $\vec{\tilde{f}}(\vec{\bar{B}}_\perp, \vec{x}_\perp,z)$ satisfies
\bea
&&i \f{\partial}{\partial z} \vec{\tilde{f}}(\vec{\bar{B}}_\perp, \vec{x}_\perp,z) \simeq \left\{ \f{1}{2\omega}\nabla_{\bar{B}}^2 -  \f{1}{2} \omega \omega_0^2 \vec{\bar{B}}_\perp^2 \right\} \vec{\tilde{f}}(\vec{\bar{B}}_\perp, \vec{x}_\perp,z)\label{equ:fz}
\eea
with
\be
\omega_0 = \f{1}{2}( 1 + i ) \left( \f{N_c \hat{q}_R}{C_R \omega}\right)^{\f{1}{2}} =  \f{1}{2}( 1 + i ) \left( \f{\hat{q}_A}{\omega}\right)^{\f{1}{2}}.
\ee
In the following, we redefine the following dimensionless variables
\bea
z=\f{z}{L}, ~~\vec{B}_\perp = \sqrt{ \hat{q}_R L }\vec{B}_\perp,~~\vec{\bar{B}}_\perp =\sqrt{ \hat{q}_R L} \vec{\bar{B}}_\perp~~ \mbox{and}~~\vec{x}_\perp = \sqrt{\hat{q}_R L} \vec{x}_\perp, \label{equ:dv}
\eea
and accordingly redefine the dimensionless gluon emission/absorption amplitudes by replacing all the variables in their expressions with the corresponding dimensionless variables. 

In terms of the above dimensionless variables, the Schr\"{o}dinger-type evolution equation (\ref{equ:fz}) is the same as that for a two-dimensional harmonic oscillator with mass $m = - \f{\omega}{\hat{q}_RL^2}$ and the angular frequency $\bar{\omega}_0 = \omega_0 L$. The amplitude $\vec{\tilde{f}}(\vec{\bar{B}}_\perp, \vec{x}_\perp,z)$ is given by
\bea
&&\vec{\tilde{f}}(\vec{\bar{B}}_\perp, \vec{x}_\perp,z) = N e^{-\gamma \vec{\bar{B}}_\perp^2} \int d^2{\vec{\bar{B}}_\perp^\prime} e^{-\alpha ( \vec{\bar{B}}_\perp^\prime - \vec{C}_\perp)^2}   \vec{\tilde{f}}( \vec{\bar{B}}_\perp^\prime, \vec{x}_\perp,0)\nn\\
&&=\f{\pi N}{\alpha} e^{-\gamma \vec{\bar{B}}_\perp^2} \left\{ \vec{D}_\perp - e^{-\alpha D_\perp^2} \left[  \f{1 - e^{\alpha D_\perp^2} ( 1 - \alpha D_\perp^2 ) }{\alpha D_\perp^2} \f{\vec{x}_\perp \cdot \vec{D}_\perp \vec{D}_\perp}{D_\perp^2} -   \f{1 - e^{\alpha D_\perp^2} }{2 \alpha D_\perp^2} \vec{x}_\perp \right]\right\}
,\label{equ:ftilde}
\eea
where
\bea
\alpha &=& \f{- i m \bar{\omega}_0 }{2 \tan\bar{\omega}_0 z}=- \f{ N_c}{4 C_R \bar{\omega}_0  \tan \bar{\omega}_0  z},~\vec{C}_\perp=\f{\vec{\bar{B}}_\perp}{\cos\bar{\omega}_0 z},\vec{D}_\perp\equiv \vec{C}_\perp + \f{\vec{x}}{2},\\
\gamma &=& \f{i}{2} m \bar{\omega}_0 \tan\bar{\omega}_0 z = \f{N_c \tan\bar{\omega}_0 z}{4 C_R \bar{\omega}_0},~N= \f{m \omega_0}{2\pi i \sin\bar{\omega}_0 z} = -\f{N_c}{4\pi C_R \bar{\omega}_0 \sin\bar{\omega}_0 z}.
\eea
Accordingly, the distribution function from radiation in the LPM regime is given by
\bea
\f{dI_{rad}}{d^2p_{R\perp}} &=& \f{\alpha_s N_c}{2 } \f{N_c}{C_R} \mbox{Re} \int_0^1 dz (1-z)  \int \f{ d^2x_\perp}{(2\pi)^2} \f{ d^2B_\perp}{(2\pi)^2} \int \f{d\omega }{\omega}  \nn\\
&\times&  \left[ \vec{\bar{f}}_{0}(\vec{B}_\perp-\vec{x}_\perp, \vec{x}_\perp)+ \vec{f}_{ab}(\vec{B}_\perp, \vec{x}_\perp) \right] \cdot  \vec{ \tilde{f}}(\vec{B}_\perp - \f{\vec{x}_\perp}{2}, \vec{x}_\perp,z) e^{-\beta x_\perp^2 -i \vec{p}_{R\perp}  \cdot \vec{x}_\perp},\label{equ:dIraddp}
\eea
where $\vec{p}_{R\perp}$ is redefined by $\vec{p}_{R\perp}=\f{\vec{p}_{R\perp}}{\sqrt{\hat{q}_RL}}$ and
\be
\beta = \f{1}{4} \left( 1 - \f{N_c}{4 C_R} z \right).\label{equ:beta}
\ee
If one first integrates out $\vec{p}_{R\perp}$ in (\ref{equ:dIraddp}), $\vec{x}_\perp$ is put to be $0$. In this case,
\be
\vec{f}_{ab}(\vec{k}_\perp,0) = - \vec{\bar{f}}_0(\vec{k}_\perp,0),~\mbox{and}~\vec{f}_{ab}(\vec{B}_\perp,0) = - \vec{\bar{f}}_0(\vec{B}_\perp,0).
\ee 
Therefore, the conservation of probability in (\ref{equ:probability}) is guaranteed.
%
%
\section{ $p_{\perp}$-broadening of high energy partons from radiation}\label{sec:results}
Now we are ready to calculate $p_\perp$-broadening from radiation. It is convenient to write $\bra p_{\perp}^2 \ket_{rad}$ in the following form
\bea 
 &&\bra p_{R \perp}^2 \ket_{rad} =- \f{\alpha_s N_c}{2 } \f{N_c}{C_R} \hat{q}_R L ~\mbox{Re}\int_0^1 dz ( 1 - z )  \int \f{ d^2x_\perp d^2 p_{R\perp}}{(2\pi)^2}  \f{d\omega}{\omega}  \f{ d^2B_\perp }{(2\pi)^2} \nn\\
&&\times \left\{ e^{-\beta x_{\perp}^2} \left[ \vec{\bar{f}}_0(\vec{B}_\perp-\vec{x}_\perp, \vec{x}_\perp)+ \vec{f}_{ab}(\vec{B}_\perp, \vec{x}_\perp) \right] \cdot \vec{\tilde{f}}(\vec{B}_\perp - \f{\vec{x}_\perp}{2}, \vec{x}_\perp,z)\right\}\nabla_{x_\perp}^2 \left( e^{ -i \vec{p}_{R\perp}  \cdot \vec{x}_\perp} \right)\nn\\
&&
=- \f{\alpha_s N_c}{2 } \f{N_c}{C_R} \hat{q}_R L ~\mbox{Re} \int_0^1 dz ( 1 - z ) \int \f{d\omega}{\omega}  \f{ d^2B_\perp }{(2\pi)^2} \nn\\
&&\times  \left.\nabla_{x_\perp}^2\left\{ e^{-\beta x_{\perp}^2} \left[ \vec{\bar{f}}_0(\vec{B}_\perp-\vec{x}_\perp, \vec{x}_\perp)+ \vec{f}_{ab}(\vec{B}_\perp, \vec{x}_\perp) \right] \cdot \vec{\tilde{f}}(\vec{B}_\perp - \f{\vec{x}_\perp}{2}, \vec{x}_\perp,z)\right\}\right|_{\vec{x}_\perp = \vec{0}}, \label{equ:pt_1_general}
\eea
where we have assumed that the expression in the curly bracket has no poles in the 2-dimensional $\vec{x}_\perp$ plane. 
 Let us define $\mathcal{K}_{rg}(\omega, z)$ and $\mathcal{K}_{vg}(\omega, z)$ respectively by
\be
\f{\partial}{\partial z}\mathcal{K}_{rg}(\omega, z) =- \f{\alpha_s N_c^2}{2 C_R}  \int \f{ d^2B_\perp }{(2\pi)^2}\left.\nabla_{x_\perp}^2\left[ e^{-\beta x_{\perp}^2} \vec{\bar{f}}_0(\vec{B}_\perp-\vec{x}_\perp, \vec{x}_\perp) \cdot \vec{\tilde{f}}(\vec{B}_\perp - \f{\vec{x}_\perp}{2}, \vec{x}_\perp,z)\right] \right|_{\vec{x}_\perp = \vec{0}}\label{equ:krgdef}
\ee
and 
\be
\f{\partial}{\partial z}\mathcal{K}_{vg}(\omega, z) =- \f{\alpha_s N_c^2}{2 C_R}  \int \f{ d^2B_\perp }{(2\pi)^2}\left.\nabla_{x_\perp}^2\left[ e^{-\beta x_{\perp}^2} \vec{f}_{ab}(\vec{B}_\perp, \vec{x}_\perp) \cdot \vec{\tilde{f}}(\vec{B}_\perp - \f{\vec{x}_\perp}{2}, \vec{x}_\perp,z)\right] \right|_{\vec{x}_\perp = \vec{0}}.\label{equ:kvgdef}
\ee
In terms of $\mathcal{K}_{rg}(\omega, z)$ and $\mathcal{K}_{vg}(\omega, z)$, $p_\perp$-broadening from radiation can be expressed as
\be
\bra p_{R \perp}^2 \ket_{rad} = \hat{q}_R L ~\mbox{Re}\int \f{ d\omega }{\omega} \left\{ \int_{z_0}^1 dz \left[ \mathcal{K}_{rg}(\omega, z) + \mathcal{K}_{vg}(\omega, z) \right] - \left[ \mathcal{K}_{rg}(\omega, z_0) + \mathcal{K}_{vg}(\omega, z_0) \right] \right\},
\ee 
with $1\gg \f{\lambda_R}{ L } \simeq z_0$. 
In the so-called LPM regime, the dominant contribution comes from the region of integration with $\hat{q}_A \lambda_R^2 \lesssim \omega \lesssim  \hat{q}_A L^2$ and $1\geq z \gtrsim \f{1}{|\omega_0 L|} \simeq \f{t_{form}}{L}$. Accordingly, one has
\be
 \sin \omega_0 L z \simeq i \cos \omega_0 L z \simeq \f{i e^{-i\omega_0 L z}} 2.
\ee
\subsubsection{$p_{\perp}$-broadening from medium-induced real gluon emission}
Inserting (\ref{equ:f0bar}) and (\ref{equ:ftilde}) into (\ref{equ:krgdef}), after some algebra we get $\mathcal{K}_{rg}(\omega, z)$ as follows
\bea
\mathcal{K}_{rg}(\omega, z) & =& \f{\alpha_s N_c}{\pi} \left\{ \f{ 2 C_R \omega_0 L }{N_c \tan( \omega_0 L z )} - \f{\omega_0 L z}{ 2\tan( \omega_0 L z )} +\f{1}{2} -  \log\left( \cos\f{ \omega_0 L z}{2} \right) \right.\nn\\
&+& \left.  \log\left( \cos ( \omega_0 L z ) \right) +\f{  \cos( \omega_0 L z ) \log\left(  -\cot^2( \omega_0 L z ) \right)}{2} \right\},\label{equ:krg}
\eea
and, therefore,
\be
-\mathcal{K}_{rg}(\omega, z_0)= - \f{\alpha_s N_c}{\pi} \left\{ \f{ 2 C_R }{N_c z_0} + \log \f{ 1 }{  \omega_0 L z_0} \right\}\label{equ:nkrg0}
\ee
with $z_0 \rightarrow 0$. Here, the first term proportional to $\f{1}{z_0}$ will be completely cancelled by the contributions from self-energy diagrams. The second term gives a double logarithmic contribution to $\bra p_\perp^2\ket$, which is parametrically the same as that in single scattering\cite{Gunion:1981qs}.

In the following, let us focus on the LPM regime. By keeping the dominant contributions from the region $1 \gtrsim z \gtrsim \f{t_{form}}{L}$ we get
\bea
\left.\omega \f{ d }{ d\omega } \bra p_{R \perp}^2 \ket_{rg} \right|_{LPM } &\simeq&\hat{q}_R L ~\mbox{Re}  \int_{\f{1}{|\omega_0 L|}}^1 dz \mathcal{K}_{rg}(\omega, z)\nn\\
&\simeq&   \f{\alpha_s N_c}{2 \pi} \hat{q}_R L \left\{ \f{ 2 C_R}{ N_c}  \sqrt{\f{\hat{q}_A L^2}{\omega}} +  1\right\}.\label{equ:odp2rgmddo}
\eea
In the above equation, the first term on the right-hand side can be easily understood according to the soft gluon spectrum in the LPM regime, which is given by \cite{Baier:1996kr}
\be
\omega \f{dI}{d\omega} \simeq \f{\alpha_s C_R}{\pi} \sqrt{\f{\hat{q}_A L^2}{\omega}}.
\ee
That is, the first term is the $p_\perp$-broadening from multiple scattering times the probability for the parton emitting a gluon. And the second term is the $p_\perp$-broadening accumulated during the gluon emission time $t_{form}$.
%
%
\subsubsection{$p_\perp$-broadening from medium-induced self-energy}
By inserting (\ref{equ:fab}) and (\ref{equ:ftilde}) into (\ref{equ:kvgdef}), we obtain
\bea
\mathcal{K}_{vg}(\omega, z) & =& \f{\alpha_s N_c}{\pi} \left\{ - \f{ 2 C_F \omega_0 L }{N_c \tan( \omega_0 L z )} + \f{\omega_0 L z}{ 2 \tan( \omega_0 L z )}  -\f{1}{2} -  \log \cos( \omega_0 L z ) \right.\nn\\
&+& \left. \log\left( 2 \sin\f{ \omega_0 L z}{2} \right) + \f{ \cos( \omega_0 L z ) \log\left(  -\cot^2( \omega_0 L z ) \right)}{2} \right\}.\label{equ:kvg}
\eea
In the limit $z=z_0\rightarrow0$,
\bea
-\mathcal{K}_{vg}(\omega, z_0) = \f{\alpha_s N_c}{\pi}\f{2C_F}{N_c z_0},\label{equ:nkvg0}
\eea
which exactly cancels the first term in (\ref{equ:nkrg0}). In the LPM regime, the dominant contribution from medium-induced parton self-energy is as follows
\bea
\left.\omega \f{ d }{ d\omega } \bra p_{R \perp}^2 \ket_{vg} \right|_{LPM } &\simeq&\hat{q}_R L ~\mbox{Re}  \int_{\f{1}{|\omega_0 L|}}^1 dz \mathcal{K}_{rg}(\omega, z)\nn\\
&\simeq&   \f{\alpha_s N_c}{2 \pi} \hat{q}_R L \left\{ -\f{ 2 C_R}{ N_c}  \sqrt{\f{\hat{q}_A L^2}{\omega}} +2 \log 2 -1\right\}.\label{equ:odp2vgmddo}
\eea
\subsubsection{$p_\perp$-broadening from radiation}
After summing over (\ref{equ:krg}), (\ref{equ:nkrg0}), (\ref{equ:kvg}) and (\ref{equ:nkvg0}), we finally obtain the the complete contribution from medium-induced gluon-emission and self-energy diagrams as follows
\bea
 \bra p_{R \perp}^2 \ket_{rad} &=&  \f{\alpha_s N_c}{\pi}  \hat{q}_R L ~\mbox{Re} \int \f{d \omega}{\omega} \left\{  \int_{z_0}^{1} d z \left[ \log\left( 2 \sin\f{ \omega_0 L z}{2} \right) + \cos( \omega_0 L z ) \log\left(  -\cot^2( \omega_0 L z ) \right)\right.\right.\nn\\& -& \left. \log\left( \cos\f{ \omega_0 L z}{2} \right) \right]
  + \left. \log\left( \omega_0 L  z_0\right) \right\},\label{equ:pt2rad_full}
\eea
which is singular in the limits $\omega\rightarrow0$ and $z_0\rightarrow0$. Therefore, in order to numerically evaluate (\ref{equ:pt2rad_full}), one needs to give an infrared cutoff $\mu$ to the $\omega$ integration and take $z_0\simeq \f{\lambda_R}{L}$. In this paper, we only discuss the parametric behavior of the above equation. In the LPM regime, one has
 \bea
\left.\omega \f{ d }{ d\omega } \bra p_{R \perp}^2 \ket_{rad} \right|_{LPM } &\simeq& \f{\alpha_s N_c \log 2}{ \pi} \hat{q}_R L .\label{equ:odp2doLPM}
\eea
%
%
\section{Discussions}\label{sec:discussions}
In this work, we calculate  $p_\perp$-broadening of a high energy parton from radiation in a QCD medium with a finite length. Both medium-induced gluon emission and self-energy diagrams are evaluated in the BDMPS formalism. We find the following parametrically large corrections to the leading order result of $p_\perp$-broadening:
\begin{itemize}
\item Double-logarithmic enhancement\\
With $z_0 L \simeq \lambda_R$, the mean free path of the parton, the last term in (\ref{equ:pt2rad_full}) is
 \bea
\left.\omega \f{ d }{ d\omega } \bra p_{R \perp}^2 \ket_{rad} \right|_{s} &\simeq& \f{\alpha_s N_c}{ 2 \pi} \hat{q}_R L \log\left( \f{\hat{q} \lambda_R^2}{2 \omega}\right) .\label{equ:odp2dos}
\eea
It has the same parametric form as the result in single scattering in (\ref{equ:idtypeI}). Therefore, this term results from incoherent gluon emission. 
\item Logarithmic enhancement\\
In the LPM regime, the contributions both from real gluon emission and parton self-energy take the following parametric from
\be
\left.\omega \f{ d }{ d\omega } \bra p_{R \perp}^2 \ket_{rad} \right|_{LPM} \simeq\f{\alpha_s N_c}{\pi} \hat{q}_R L.
\ee
The physical interpretation is as follows: the emitted (real/virtual) gluon with energy $\omega$ is finally freed from the high energy parton after the formation time $t_{form}$, which is defined by
\be
t_{form} \simeq \f{ \omega }{ k_\perp^2 } \simeq \f{ \omega }{ \hat{q}_A t_{form} },
\ee
that is,
\be
t_{form} \simeq \sqrt{\f{\omega}{ \hat{q}_A} }.
\ee
During such a time period, the parton picks up a typical transverse momentum broadening equal to $\hat{q}_R t_{form}$. Since we are only interested in the case with the length of the medium less than the critical length, that is,  \be L\leq L_{cr}\equiv \sqrt{ \f{E}{ \hat{q}_A} },\ee
we need only to account the contributions from $\left( \f{ L }{ t_{form} } \right)$ coherent regions. Within each coherent region, all the $\left( \f{ \mathcal{N} t_{form}}{L} \right)$ constituent particles behave like one single scatterer. Here, we assume that there are totally $\mathcal{N}$ scatterers in the medium. As a result, the total transverse broadening of the parton is given by
\be
\left.\omega \f{ d }{ d\omega } \bra p_{R \perp}^2 \ket_{rad} \right|_{LPM} \simeq \f{\alpha_s N_c}{\pi} \left( \f{L}{t_{form}}\right)  \hat{q_R} t_{form}=\f{\alpha_s N_c}{\pi} \hat{q}_R L,
\ee
where we have taken into account the probability for the whole process by putting in the factor $\left( \f{\alpha_s N_c}{\pi} \right)$.
\end{itemize}
Therefore, in a medium with $\hat{q}_A L^2\gg \mu\gtrsim \Lambda_{QCD}$ we expect a parametrically large correction from radiation to the medium-induced $p_\perp$-broadening in perturbative QCD.
%
%
\section*{Acknowledgements}
The author is greatly indebted to Prof. A. H. Mueller for numerous illuminating discussions and comments. Inspiring discussions with S. Caron-Huot are sincerely appreciated. The author would also like to thank R. Baier, M. Laine, B.-Q. Ma, M. Martinez-Guerrero, Y. Mehtar-Tani, P. Romatschke, C. A. Salgado and A. Vuorinen for useful discussions and/or comments.
This work was supported by the Helmholtz International Center for FAIR within the framework of the LOEWE program launched by the State of Hesse and partially by the Humboldt foundation through its Sofja Kovalevskaja program.




\begin{thebibliography}{99}


\bibitem{Baier:1996sk}
  R.~Baier, Y.~L.~Dokshitzer, A.~H.~Mueller, S.~Peigne and D.~Schiff,
  Nucl.\ Phys.\  B {\bf 484}, 265 (1997)
  [arXiv:hep-ph/9608322].

\bibitem{Baier:2002tc}
  R.~Baier,
  Nucl.\ Phys.\  A {\bf 715}, 209 (2003)
  [arXiv:hep-ph/0209038].
    
\bibitem{Baier:1996kr}
  R.~Baier, Y.~L.~Dokshitzer, A.~H.~Mueller, S.~Peigne and D.~Schiff,
  Nucl.\ Phys.\  B {\bf 483}, 291 (1997)
  [arXiv:hep-ph/9607355].
    
\bibitem{Baier:1998kq}
  R.~Baier, Y.~L.~Dokshitzer, A.~H.~Mueller and D.~Schiff,
  Nucl.\ Phys.\  B {\bf 531}, 403 (1998)
  [arXiv:hep-ph/9804212].
  
\bibitem{Wu:2010ze}
  B.~Wu and B.~Q.~Ma,
  Nucl.\ Phys.\  A {\bf 848}, 230 (2010)
  [arXiv:1003.1692 [hep-ph]].

\bibitem{Eskola:2004cr}
  K.~J.~Eskola, H.~Honkanen, C.~A.~Salgado and U.~A.~Wiedemann,
  Nucl.\ Phys.\  A {\bf 747}, 511 (2005)
  [arXiv:hep-ph/0406319].
  
\bibitem{Baier:2006fr}
  R.~Baier and D.~Schiff,
  JHEP {\bf 0609}, 059 (2006)
  [arXiv:hep-ph/0605183].
  
\bibitem{Dominguez:2008be}
  F.~Dominguez and B.~Wu,
  Nucl.\ Phys.\  A {\bf 818}, 246 (2009)
  [arXiv:0811.1058 [hep-ph]].

\bibitem{Dominguez:2008aa}
  F.~Dominguez, C.~Marquet and B.~Wu,
  Nucl.\ Phys.\  A {\bf 823}, 99 (2009)
  [arXiv:0812.3878 [nucl-th]].
 
\bibitem{CaronHuot:2008}
  S.~Caron-Huot,
  Phys.\ Rev.\  D {\bf 79}, 065039 (2009)
  [arXiv:0811.1603 [hep-ph]].
 
\bibitem{Dominguez:2008vd}
  F.~Dominguez, C.~Marquet, A.~H.~Mueller, B.~Wu and B.~W.~Xiao,
  Nucl.\ Phys.\  A {\bf 811}, 197 (2008)
  [arXiv:0803.3234 [nucl-th]].

\bibitem{Gunion:1981qs}
  J.~F.~Gunion and G.~Bertsch,
  Phys.\ Rev.\  D {\bf 25}, 746 (1982).

 \bibitem{Zakharov:1996fv}
  B.~G.~Zakharov,
  JETP Lett.\  {\bf 63} (1996) 952
  [arXiv:hep-ph/9607440];
  B.~G.~Zakharov,
  JETP Lett.\  {\bf 65} (1997) 615
  [arXiv:hep-ph/9704255].
  
\bibitem{glv}
  M.~Gyulassy, P.~Levai and I.~Vitev,
  Nucl.\ Phys.\  B {\bf 594}, 371 (2001)
  [arXiv:nucl-th/0006010];
  M.~Gyulassy, P.~Levai and I.~Vitev,
  Phys.\ Lett.\  B {\bf 538}, 282 (2002)
  [arXiv:nucl-th/0112071].
  
\bibitem{amy}
  P.~B.~Arnold, G.~D.~Moore and L.~G.~Yaffe,
  JHEP {\bf 0206}, 030 (2002)
  [arXiv:hep-ph/0204343].
  
\bibitem{ksw}
A.~Kovner and U.~ A.~ Wiedemann, arXiv:hep-ph/030415;
 J.~Casalderrey-Solana and C.~A.~Salgado,
  Acta Phys.\ Polon.\  B {\bf 38}, 3731 (2007)
  [arXiv:0712.3443 [hep-ph]].

\bibitem{Wiedemann:2009sh}
  U.~A.~Wiedemann,
  arXiv:0908.2306 [hep-ph].
 
  \end{thebibliography}
\end{document}